Examining the Impact of Source-product Congruence and Sponsorship Disclosure on

the Communicative Effectiveness of Instagram Influencers

Yi Xin Lim
Undergrad Student

Weiyu Zhang*
Associate Professor, PhD

Department of Communication and New Media
Faculty of Arts and Social Sciences
National University of Singapore

*Corresponding author can be reached at Block AS6, Room 03-24, 11 Computing Drive,
Singapore 117416. Tel: (65) 6516 8156, Fax: (65) 6779 4911, Email:
weiyu.zhang@nus.edu.sg

**Acknowledgement**
This work was developed as part of Yi Xin's honor thesis. Weiyu was the thesis supervisor.
The work did not receive any funding.

**Abstract**

Guided by the Persuasion Knowledge Model and the Attribution Theory, this study investigates the perceived source expertise-product attribute congruence and sponsorship disclosure as pertinent factors affecting the communicative effectiveness of influencers. Instagram, with an immense influencer market value projected at $2.3 billion in 2020, was chosen as the platform context. The study utilised a 2 (source expertise) x2 (product category) x2 (sponsorship disclosure) experiment to examine the roles of source-product congruence and sponsorship disclosure in affecting consumer's perception of extrinsic and intrinsic source motives, consumer resistance and ultimately, advertising effectiveness. Results revealed that the presence of a sponsorship disclosure generated stronger perceptions of extrinsic source motives but did not impact consumer resistance and advertising effectiveness, indicating that the activation of consumer's conceptual persuasion knowledge may not necessarily affect attitudinal persuasion knowledge. Source-product congruence, on the other hand, had main impacts on intrinsic motives, consumer resistance and ad effectiveness. In addition, hierarchical multiple regressions found that source-product congruence triggers a multi-stage process where consumer's perception of intrinsic source motives mediates consumer resistance which subsequently, mediates the relationship between source-product congruence and ad effectiveness.

The Impact of Source-product Congruence and Sponsorship Disclosure on the

Communicative Effectiveness of Instagram Influencers

Social media today offers new possibilities for organisations to reach audiences. The rise of social media channels, along with a new form of sources, now commonly known as social media influencers (SMIs), have resulted in a new paradigm for source effects study. SMIs are commonly understood as a group of people who have achieved "instafame" through their online social activity (Yates, 2014). These SMIs have been recognised by Freberg and colleagues (2011) as representing a "new type of independent third party endorser who shape audience attitudes through blogs, tweets, and the use of other social media." (p.90). SMIs are also commonly known as micro-celebrities on social networking sites (SNS) (Chae, 2018). Contrary to mainstream media celebrities who appeal to the general population and garner large-scale followings, microcelebrity involves fame amongst a small, niche community and the curation of an authentic persona to their followers (Marwick, 2013).

Social media posts featuring brand-related information created by SMIs are electronic word of mouth (eWOM) messages. These messages are defined as "any positive or negative statement made by potential, actual, or former customers about a product or company, which is made available to a multitude of people and institutions via the Internet" (Hennig-Thurau et al., 2004, p. 39). These eWOM messages are perceived to be honest and unbiased consumer opinions (Chatterjee, 2011) and have therefore been found to be effective in influencing product evaluations (Lee & Youn, 2009), product decisions (Huang & Chen, 2006), brand image (Sandes & Urdan, 2013), brand attitudes (Lee, Park, & Han, 2008; Lee, Rodgers, & Kim, 2009), usefulness presence (Kumar & Benbasat, 2006), social presence (Kumar & Benbasat, 2006), purchase intentions (Bickart & Schindler, 2001; Park & Lee, 2008; Park, Lee, & Han, 2007; Sandes & Urdan, 2013; Xia & Bechwati, 2008) and product sales (Babić et al., 2016; Chevalier & Mayzlin, 2006; Liu, 2006). Recognising the influence of eWOM, marketers have started to increasingly tap on influencer marketing – the commercial activity of organisations employing social media personalities to advertise products.

While social media and influencer marketing now offer a less obtrusive communication channel for marketers to tap on, many cumulative events over the past few years have caused a taint in the business of influencer marketing. The FYRE festival catastrophe, where supermodel influencers such as Bella Hadid publicised an extravagant image of a festival that turned out to be a spectacular flop, the Logan Paul suicide forest saga where famous YouTube star Logan Paul posted a dead body of a suicide victim in Japan, and the recent exposé of a vegan YouTuber Yovana Mendoza caught on camera eating fish, alongside revelations of various corruptive practices such as buying of fake followers, likes and undeclared sponsored content have led to widespread speculation of the impending doom of influencer marketing (Elmhirst, 2019).

Closer to the context of Country A, the recent engagement of influencers as sources of government communication campaigns have received critique and skepticism. For instance, during Country A's Budget 2018, the Ministry of Finance (MOF) employed over 50 SMIs to promote the Budget 2018 process and to encourage citizens to provide feedback (Seow, 2018). Despite well intended motivations to promote youth engagement, the campaign met with waves of backlash that received coverage in mainstream news media. Critics argued that the influencers provided only superficial awareness that feedback channels exist (Seow, 2018), and the mismatch between the low financial expertise of the SMIs and the topic of budget resulted in posts that seemed out-of-place and ingenuine (Zannia, 2018).

Despite various events that have shifted the spotlight over the artiface of the internet and influencer ecosystem, the influencer market value on Instagram alone is projected to still rise from $800 million in 2017 to $2.3 billion in 2020 (Elmhirst, 2019). In an attempt to regain trust, the industry has now been trying to move towards a new phase of authenticity. Today, factors affecting the authenticity, or at least the appearance of authenticity, is vital amidst the increasingly informed audience (Elmhirst, 2019). In light of these events, this study investigates the perceived match between source and product as well as sponsorship disclosure as two pertinent factors affecting consumer responses towards influencer's sponsored content. Moreover, this study seeks to unravel the plausible process on how both studied factors – sponsorship disclosure and source-product congruence – affects consumers responses to the sponsored content.

## Literature Review

### Source Effects Literature

Source effects literature comprises of four main streams where research has been done: (1) the source credibility model, (2) the source attractiveness model, (3) the congruence or match-up model and (4) the meaning transfer model (MTM) (Erdogan, 1999). The first two models are part of early literature on source effects which espoused on the effects of positive source characteristics on audience's attitudes regardless of topic or product variations. Extant research culminated in the identification of several prominent source attributes that wield far-reaching impact on audience's receptiveness and attitudes. This includes attributes such as source credibility or expertise, source attractiveness and source similarity. Kelman (1961), for instance, suggested that the three identified source characteristics, credibility, attractiveness and power to reward or punish, may lead to attitude change via three psychological modes, (1) internalisation, (2) identification and (3) compliance respectively. The widely adopted Source Credibility Model (Hovland, 1953; Ohanian, 1990) and the Source Attractiveness Model (McGuire, 1985) posit that certain desirable source attributes can positively influence audience's reception of a conveyed message (Erdogan, 1999).

The source credibility model focuses on the perceived credibility of the source by a receiver (Cheung & Thadani, 2012). The model is centered on two key dimensions – expertise and trustworthiness – which makes up the multi-dimensional concept of credibility perception (McCracken, 1989; Ohanian, 1990). Expertise is defined as "the extent to which a communicator is perceived to be a source of valid assertions" (Hovland, 1953, p. 21). In the context of a social media platform like Instagram, social cues such as number of followers and "likes" reflect the perceived expertise of a source that may contribute to a SMI's persuasiveness (Martensen, Brockenhuus-Schack, & Zahid, 2018). Trustworthiness, the other key dimension of the source-credibility model, is defined as "the degree of confidence in the communicator's intent to communicate the assertions he considers most valid" (Hovland, 1953, p. 21). SMIs and citizen influencers are perceived to be trustworthy relative to other sources due to the inclusion of personal opinions and expressions in their online content (Martensen, Brockenhuus-Schack, & Zahid, 2018). Source expertise and trustworthiness have been shown to be positively related with consumer's attitudes (Johnson, Torcivia, & Poprick, 1968), compliance (Crisci & Kassinove, 1973), behavioural intentions and actual behaviours (Senecal & Nantel, 2004).

McGuire (1969) introduced the source attractiveness model which suggests that source attractiveness, a multi-dimensional concept made up of factors such as likability, similarity and familiarity, is also a factor that has attitudinal impact on receivers (McGuire, 1985). Source attractiveness reflects the degree in which the audiences identifies with the source and views the source as a referent other (Antil et al., 2012). The model posits that there exist a positive relationship between perceived attractiveness and persuasiveness of the source (Ohanian, 1990). Previous research have found that perceived similarity plays an essential part in affecting SMI's persuasiveness, since it facilitates easy identification with sources of messages (Martensen, Brockenhuus-Schack, & Zahid, 2018). Thus, source attractiveness should be seen to possess immense impact on a source's persuasive power and their influence on brand attitudes.

However, the presence of unsuccessful source dominant marketing campaigns illustrates the rigidity of the traditional source effects model, which assumes a constant relationship between source characteristics and impact. A review revealed that only 53% of the cited effects of source characteristics on consumer's attitudes were statistically significant, suggesting the inability of traditional source effects theories to account fully for the data (Wilson & Sherrell, 1993). As such, there is a need to examine other factors that impact consumers evaluations of source-dominated communication materials, among which this study examines product-source congruence and disclosure of sponsorship.

**Source-Product Congruence**

Contrary to constant source effects proposed in traditional source effects models, other researchers propose that positive source characteristics may only enhance product and ad based evaluations if source characteristics match up to product attributes (Baker & Churchill, 1977; Friedman and Friedman, 1979; Kahle & Homer, 1985; Kamins, 1990; Peterson & Kerin, 1977). The match-up hypothesis posits that congruence between source and product chrateristics in advertisements impacts communication effectiveness and brand evaluations. Friedman & Friedman (1979) conducted an experiment that manipulated endorser type (celebrity, expert, typical consumer and no-endorsement control) and product category (costume jewelry, vacuum cleaner and box of cookies). Consistent with the match-up hypothesis, results revealed that highest evaluations were achieved by the pairing of celebrity with costume jewelry, the expert with vacuum cleaner, and the typical consumer with cookies treatment.

Other match-up studies focused on the match between source and product based on physical attractiveness (Kahle & Homer, 1985; Kamins, 1990). These studies predict that the effectiveness of source attractiveness is enhanced when matched with products that are used to enhance one's attractiveness. Source expertise, amongst other source characteristics (e.g. source attractiveness, trustworthiness, expertise), have been found to be most closely related to purchase intentions of the endorsed product (Ohanian, 1990) Till and Busler's (2000) study found that an athlete was the most effective in increasing brand attitudes when matched with an energy bar, indicating the appropriateness of expertise as a match-up dimension. Studies have found that seductive and attractive female models were more favoured and effective for beauty products such as body oil (Peterson & Kerin, 1977), perfume (Baker & Churchill, 1977) or razors (Kahle & Homer, 1985) as compared to products such as ratchet sets (Peterson & Kerin, 1977), books (Caballero & Solomon, 1984) or computers (Kamins, 1990).

**Sponsorship Disclosure**

Sponsored influencer content highly resemble regular posts of SMIs sharing their daily life, their interests and the products they have been enjoying, making it increasingly difficult for consumers to differentiate commercial posts from posts with non-commercial intent (Shrum, 2012; Lueck, 2015). To ensure proper discernment and fairness in commercial practices, the Federal Trade Commission (FTC) has required the presence of a clear and salient disclosure declaring the commercial and paid relationship in an influencer's sponsored post in December 2015. Following which, the Advertising Standards Authority of Country A (ASAS) on 29 August 2016 issued guidelines on Interactive Marketing Communication and Social Media, stipulating amongst other guidelines, that disclosures in marketing communication and sponsored content should be clear and prominent (Advertising Standards Authority of Country A (ASAS, 2016). Despite the policy advances, scant research has been done on the effects of sponsorship disclosure. Moreover, we still lack understanding about how source characteristics such as congruence and disclosure influence consumers' perceptions about the sources and eventually lead to their evaluations of ads and products. We propose that we need to examine both conceptual and attitudinal persuasion knowledge.

**Conceptual Persuasion Knowledge and Source Motives**

Persuasion knowledge helps individuals identify, interpret and cope with various advertising tactics (Friestad & Wright, 1994; Hibbert et al., 2007). The model states that audiences need to be cognisant of a persuasion attempt before they can activate the cognitive process of retrieving their persuasion knowledge in order to cope with the attempt. This cognitive dimension, known as conceptual persuasion knowledge, consists of the ability to recognise advertising, understand advertisement's persuasive intent, selling intent and coping tactics (Rozendaal et al., 2011).

Attribution Theory has also been widely adopted in the understanding of how receivers infer source's motivations in communicating a message (Choi & Rifon, 2012; Kamins, 1990; Kelley, 1973). The theory posits that consumers cognitively infer two types of attribution; receivers can either attribute a communicator's action with an extrinsic/ situational factor (e.g. environment/ situation based) or an intrinsic/ dispositional factor (e.g. personal characteristics). Jiang (2018) consolidated the attribution theory, persuasion knowledge model and eWOM literature to conceptualise a scale that captures six distinct motives an influencer can possess when sharing a post on social media. Extrinsic motives included (1) money motives — influencer is seen as having a material connection with an corporate organisation; (2) image motives — influencers post in order to attain their desired self-image and/ or seek positive evaluations; and (3) selling motives — influencer wishes to promote the product. Intrinsic motives, on the other hand, included (4) love motives — influencer's post is motivated by a genuine liking and satisfaction of the product, (5) sharing motives — influencer is driven by a desire to maintain bonds and relationships on the social platform and (6) helping motives —influencer wishes to aid others in their purchase decisions. The attribution process is subject to the discounting principle (Jones & Davis, 1965; Kelley, 1973), which proposes that "the role of a given cause in producing an effect is discounted if other plausible causes are also presented" (Kelley, 1971, p.8). As such, consumers tend to discount a product recommendation when they recognise that a recommendation is caused by motivations other than product attributes (Brandt, Vonk, & van Knippenberg, 2011).

Although limited research has been done on the effects of social media influencers, it has been found that there is a general predisposition for audiences to attribute celebrity endorsements to intrinsic motivations such as a genuine liking of the product as opposed to a financially motivated extrinsic motive (Atkin & Block, 1983; Boerman, Willemsen & Van Der Aa, 2017). This trust in the purity of celebrities' motives have been explained by the correspondent inference bias rooted in the attribution theory. Influencers typically embed endorsed products into personal stories, depicting their personal use and enjoyment of the product (Lueck, 2015). Studies have found that congruence between sponsor and its cause has a significant impact on perception of greater altruistic motives that in turn generated stronger positive perceptions of source credibility and consumer attitudes (Rifon et al., 2004). On the contrary, we expect that if an influencer were to share a product whose attributes are distinctly different from his/her image, there is a higher tendency for the communication content to appear less genuine and believable. In consideration of the above discussion, the study proposes the following hypothesis:

H1a: Source-product incongruence will create perceptions of greater extrinsic motives than source-product congruence.

H1b: Source-product congruence will create perceptions of greater intrinsic motives than source-product incongruence.

The inclusion of a sponsorship disclosure is intended to heighten awareness towards the commercial nature of a piece of sponsored content. Accordingly, viewers who have comprehended the information will recognise the persuasive and selling intent of the sponsored Instagram post (D'Astous & Chartier, 2000). Sponsorship disclosures have been found to reduce the credibility of messages (Wojdynski, 2016) and increase skepticism (Boerman, van Reijmersdal, & Neijens, 2012). Empirical evidence have shown that a disclosure that meets FTC's issued endorsement guidelines (Federal Trade Commission, 2013) activates persuasion knowledge (Boerman, Willemsen & Van Der Aa, 2017). Taking the persuasion knowledge model and attribution theory together, a sponsorship disclosure will activate cognitive persuasion knowledge. Consequently, we expect that consumers who are cognisant of the fact that the source had received some form of commercial incentive for the communication content will tend to attribute the post to extrinsic motives rather than intrinsic motives.

H2a: Sponsorship disclosure will create perceptions of greater extrinsic motives than sponsorship non-disclosure.

H2b: Sponsorship non-disclosure will create perceptions of greater intrinsic motives than sponsorship disclosure.

**Attitudinal Persuasion Knowledge and Advertising Effectiveness**

The change of meaning principle posited in the persuasion knowledge model is useful to understand and examine consumer's resistance against persuasion attempts (Friestad & Wright, 1994). The principle states that when consumers gain knowledge of existing persuasion strategies or tactics (e.g. recognition of sponsorship), a "detachment effect" follows where consumers disengage from the persuasion interaction and discount what the influence agent says (Friestad & Wright, 1994), mitigating the message persuasiveness. Following the perception of source's ulterior motives, sponsored posts will take on a change of meaning and consumers are likely to demonstrate resistance towards the influencer's messages and persuasion. In line with this proposition, Hass and Grady (1975) have found that forewarning of communicator's persuasive intent results in higher resistance toward communication messages. Various studies have similarly provided evidence that consumers who have greater persuasion knowledge are more likely to resist persuasion (Bambauer-Sachse & Mangold, 2013; Campbell & Kirmani, 2000; Morales, 2005; Wojdynski, 2016). Sponsorship disclosure and poor congruence have been found to yield more cognitive evaluation and greater resistance to the sponsored message (Cain, 2011; Menon & Kahn, 2003).

Audiences may also use their attitudinal persuasion knowledge to interpret a message (Rozendaal et al., 2011). This comprises of various attitudinal processes that aid individuals in coping with advertising messages such as development of critical attitudes. An activation of conceptual persuasion knowledge have also been found to lead to attitudinal persuasion knowledge where the awareness of a persuasion intent has led to increased critical feelings toward the advertisement (Boerman, van Reijmersdal, & Neijens, 2012). This phenomenon can be explained by the reactance theory (Brehm, 1966). The theory posits that people desire to maintain their individual freedom and autonomy. As such, when met with situations where they feel influenced or manipulated, they seek to oppose the persuasive appeal.

The discounting principle under attribution theory further explains how resistance arises as a result of attributions to extrinsic motives. The principle predicts that the attribution to the extrinsic motives such as money motives or image motives will discount the persuasiveness of the message and invite defensiveness against the message. Yoon and colleagues (2006) have found that when a consumer attributes the company's corporate social responsibility (CSR) activities to insincere motives (e.g. money), CSR efforts can actually backfire, leaving the company with a more negative image than before. The discounting principle also helps to explain consumer responses to other non-selling and non-persuasive intents, such as genuine love of the product, or pure sharing or helping intention. The principle suggests that taking attributions to intrinsic reasons may reduce consumer's resistance. For example, Yoon and colleagues (2006) have found that when a consumer attributes the company's CSR activities to sincere motives (intrinsic), the company image is actually improved. Furthermore, Sørum, Grape, and Silvera (2003) found that making attributions to intrinsic reasons such as believing the source truly likes the product (love motives) had a positive effect on evaluations of the advertisement. Based on the change of meaning principle, the discounting principle and previous studies, perceptions of source motives as extrinsic and intrinsic will result in varying levels of resistance towards communication material.

H3: Source-product incongruence will create greater consumer resistance than source-product congruence.

H4: Sponsorship disclosure will create greater consumer resistance than sponsorship non-disclosure.

It has also been demonstrated that when resistance of a persuasive message occurs, attitudes of audiences are likely to become more unfavourable (Tormala & Petty; 2002, as seen in Boerman, van Reijmersadal & Neijens, 2012). That is, when individuals recognise that they have resisted persuasion successfully, they are likely to draw corresponding inferences about their attitudes, therefore increasing their attitude certainty (Tormala & Petty; 2002). In line with this observation, studies have found a positive relationship between activation of persuasion knowledge and lowered favourable brand attitudes (Lee, 2010; Wei, Fischer, & Main, 2008). Sponsorship disclosure has been found to activate resistance due to credibility judgements that consequently come to affect the perceived credibility and effectiveness of advertisements (Byrne et al., 2012). Likewise, congruence wields crucial impact on consumer's predisposition toward advertisements, which in turn influences brand beliefs and purchase intentions (Fleck, Korchia & Le Roy; 2012). Taking these prior studies, we draw the hypothesis that source-product incongruence and sponsorship disclosure will both negatively impact the effectiveness of advertisements.

H5: Source-product congruence will create greater ad effectiveness than source-product incongruence.

H6: Sponsorship non-disclosure will create greater ad effectiveness than sponsorship disclosure.

**The Plausible Process of Influence**

Finally, this study also seeks to explore a step by step process in terms of the relationship from source-product congruence/sponsorship disclosure, to motive perceptions, to consumer resistance, and finally to perceived ad effectiveness. It appears chronically plausible given that detection of source's motives after seeing the manipulated messages affects consumer resistance as suggested by the persuasion knowledge model. The affected consumer resistance may subsequently lead to evaluations of ad effectiveness, according to the predictions implied by xxx theory. Therefore, we ask a research question to investigate possible mediating effects on consumer resistance and ad effectiveness.

RQ: Is there any mediation between source-product congruence/sponsorship disclosure, motive perceptions, consumer resistance, and perceived ad effectiveness?

**Method**

**Preliminary Study**

A pre-test study was administered to facilitate the planning and conduct of the main experiment. As suggested by previous literature, pilot or preliminary studies should have sample sizes of between 10-30 participants (Isaac & Michael, 1995). As such, 20 participants were recruited for the pre-test. Participants were between the ages of 18-34 with similar demographics as participants recruited for the main study. Pre-test study participants were asked to indicate their perceived gender image, familiarity and likeability of ten gender neutral names and ten fictitious brand names. Lastly, participants also indicated their levels of liking for ten images featuring either a health or finance related image. According to the pretest results, we chose the influencer name perceived to be the most gender-neutral and with a low familiarity but high likability rating. Similarly, we chose brand names for the two respective products rated as the most likable with a relatively low familiarity. Lastly, product images with similar scores of likeability were chosen to be used in the sponsored post stimuli.

**Manipulation**

**Communication Materials.** A mock Instagram post was designed after influencer name and product picture was selected according to results from the pretest. The sponsored content is adapted from real Instagram sponsored posts on nutrition and finance products. The brand information was integrated into the content to mimic the sponsored post. The study only contained positive product information with the intention to promote the brand and wording was kept as similar as possible between the content of the two products.

**Source Expertise Area Manipulation.** To manipulate knowledge of influencer's relevant expertise, participants were tasked to read a short profile of the influencer prior to reading the sponsored post. However, instead of priming audiences to perceive differences in influencer's credibility through painting one as a qualified and the other as an amateur, the profile primed audiences by showing one influencer as having expertise in nutrition and the other in finance.

The profiles stated that both influencers with expertise in separate fields were credible with a relevant advanced degree or years of working experiences with 200,000 followers. The number of the followers was derived as an average of the number of followers the 20 top influencers in Country A had in 2019 (MediaOne, 2018). This following number falls within the classification of a top influencer (Pick, 2018). The profile also stated the level of trustworthiness by indicating the sincerity of the influencer. Other character traits proven effective in predicting credibility such as down to earth and caring were also stated (Rifon, Jiang & Kim, 2016).

**Sponsorship Disclosure Manipulation.** This study designed sponsorship disclosure based on the ASAS's recommended forms of clear and prominent disclosure as well as observed real practices to enhance ecological validity (ASAS, 2016). Along with the previous findings on characteristics of effective disclosure (e.g. Boerman, van Reijmersadal & Neijens, 2014, 2015; Wojdynski, 2016) and ASAS's recommendations for written disclosures, this study uses clear and simple disclosure language, and places multiple disclosures at the beginning and at the end of the caption of the sponsored Instagram still image to ensure that the disclosure is salient. Specifically, this study uses one sentence to disclose the sponsorship at the beginning of the post, "So excited to partner with @ [brand name]," and "#sponsored" at the end of the post. In the disclosure condition, participants read a disclosure at the beginning of the article, stating "partner with [brand]" and at the end with "#sponsored." "@ [brand name]" and "#sponsored" were in the hyperlink blue color.

## Sample

A priori power analysis using G*Power with an ANCOVA main and interaction effects is used to determine the sample size. Results suggest that a sample size of at least 333 is required for the main study with an effect size of f=.21 (as in Campbell & Kirmani, 2000), power = .80 and at a significance level of .05. Adults aged 18-34, who are active users of Instagram, were the target population of the study. This age group makes up a large majority of Instagram users (61%) in Country A (Kowalczyk, 2017). 336 participants with 102 males and 232 females between the ages of 18 and 33 were recruited for the online experiment and were told that the study concerns the effectiveness of influencer marketing. The mean age of participants was 22.9 years ($SD = 2.4$).

**Procedure**

The study employed 2 (area of expertise; nutrition vs. finance) by 2 (sponsorship disclosure; disclosure vs non-disclosure) by 2 (product; protein supplement vs. finance website) factorial design. Participants were randomly assigned to one of the eight experimental conditions. There were no significant differences between the values of any of the demographic characteristics for the eight experimental conditions, indicating the validity of the random assignment.

Half of the participants were randomly assigned to read a short introduction of the influencer priming their expertise to be in the area of nutrition, while the other half read the introduction intended to prime the influencer's perceived expertise to be in the area of finance. Thereafter, participants from each expertise priming condition were exposed to an Instagram post which included a picture and message featuring prominently either a protein supplement or the finance website. Half of the participants saw a clear sponsorship disclosure, while the remaining half saw an Instagram posts that did not contain any disclosure of sponsorship.

After reading the sponsored post, participants were asked to answer a questionnaire containing the dependent measures. Participants responded to items in accordance, measuring the strength of their inferences about influencer motives, their resistance toward the posts and their ad evaluations. Following which, they answered control measures of their involvement in the sponsored content, their need for cognition, advertising skepticism and perceived appropriateness of sponsored content. They then answered manipulation check questions and lastly, they reported their demographic information.

**Dependent Measures**

*Influencer motives* were measured with the scale developed by Jiang (2018) in a five-point Likert scale, 1=Strongly Disagree, 5=Strongly Agree. Both extrinsic and intrinsic motives were measured using the Extrinsic Motives scale ($M = 3.66$, $SD = 0.57$, Cronbach's $α = .858$) which included money, selling and image motives, and the Intrinsic Motives scale ($M = 3.43$, $SD = 0.66$, Cronbach's $α = .909$) which measured love, sharing as well as helping motives.

*Consumer Resistance* was measured with seven statements adapted from van Reijmersdal et al. (2016) on a five-point Likert type scale, 1=Strongly Disagree, 5=Strongly Agree: "While reading the Instagram post, I contested/refuted /doubted /countered the information in the content;" "While reading the Instagram post, I felt angry/irritated/annoyed.". The scale possessed high internal consistency ($M = 2.62$, $SD = 0.79$, Cronbach's $α = .904$).

*Ad Effectiveness* was measured using a 10-item 5- point semantic differential scale adopted from Baker & Churchill (1977). The items have been used to measure consumer's cognitive, affective and conative advertising evaluations ($M = 2.84$, $SD = 0.58$, Cronbach's $α = .823$).

**Control Variables**

*Participants' Involvement* in the sponsored post was measured on a 10-item five-point semantic scale ($M = 2.62$, $SD = 0.76$, Cronbach's $α = .944$) adopted from Zaichkowsky (1994). *Need for Cognition* was measured on a 18-item five-point Likert scale, 1=strongly disagree, 5=strongly agree ($M = 3.06$, $SD = 0.48$, Cronbach's $α = .844$) adopted from Cacioppo & Petty (1982) and Cacioppo, Petty & Kao (1984). The scale measures the tendency in which an individual prefers to engage in effortful thinking and cognitive processes. *Advertising Skepticism* was measured on a 9-item five-point Likert scale, 1=strongly disagree, 5=strongly agree ($M = 3.51$, $SD = 0.64$, Cronbach's $α = .888$) adopted from Obermiller & Spangenberg (1998). The construct measures the degree to which an individual perceives advertisements to be truthful, believable and useful. *Perceived appropriateness of sponsored content* was measured with three adapted items from Nelson et al (2009) and Yoo (2009), was measured with three adapted items on a five-point Likert scale, 1=strongly disagree, 5=strongly agree ($M = 3.68$, $SD = 0.69$, Cronbach's $\alpha = .704$)

## Results

**Manipulation Checks**

To examine the effectiveness of experimental manipulations, participants were asked to rate their perceived congruence between the image of the influencer to the product on a 5-point semantic differential scale with the end points of "compatible/ not compatible," "relevant/ irrelevant," "good fit/ bad fit," and "good match/ bad match" (Kamins, 1994; Choi & Rifon, 2012). Independent sample T-tests showed that participants in the source-product congruence condition rated higher perceived congruence ($M = 3.98$, $SD = 1.01$) than participants in the source-product incongruence condition ($M = 2.14$, $SD = 1.11$, $t(331) = 15.91$, $p < .001$).

To measure if the sponsorship disclosure was successfully recognized by participants, they were asked to answer a single-item measure using a 5- point Likert scale, to "indicate the extent in which you thought the Instagram post was advertising" (Boerman, van Reijmersadal & Neijens, 2012; Evans, Phua & Jun, 2017). Independent sample T-tests found a significant difference in recall of sponsorship disclosure between the non-disclosure condition ($M = 2.74$, $SD = .91$) and the disclosure condition ($M = 3.97$, $SD = .84$, $t(332) = 12.84$, $p < .001$). There was also a significant difference between the disclosure and non-disclosure conditions in perception of the Instagram post as advertising, such that the participants who viewed the post with a sponsorship disclosure ($M = 3.80$, $SD = .98$) showed higher perception of the Instagram post as advertising as compared to the non-disclosure condition ($M = 2.47$, $SD = 1.06$, $t(332) = 11.96$, $p < .001$).

**Main and Interaction Effects**

Two-way ANCOVAs were conducted, controlling for consumer involvement, need for cognition, advertising skepticism and perceived sponsorship appropriateness, to examine the main and interaction effect of source-product congruence and disclosure of sponsorship on perception of source motives, consumer resistance and ad effectiveness. The results indicated no significant main effect of source-product congruence on extrinsic motives ($F(1, 328) = .057$, $p = .811$, partial $\eta^2 = .001$). However, a significant main effect of sponsorship disclosure on extrinsic motives ($F(1, 328) = 14.48$, $p < .001$, partial $\eta^2 = .042$) was found. There were no significant interaction effects between source-product congruence and sponsorship disclosure on perception of extrinsic motives, $F(1, 328) = .658$, $p = .418$, partial $\eta^2 = .002$. As such, H1a is rejected while H2a is supported.

A significant main effect for source-product congruence on perceptions of source's intrinsic motives ($F(1, 328) = 9.84$, $p = .0019$, partial $\eta^2 = .029$) was found. There is no significant main effect for sponsorship disclosure on intrinsic motives ($F(1, 328) = 1.59$, $p = .208$, partial $\eta^2 = .005$). No interaction effect was observed between source-product congruence and sponsorship disclosure on intrinsic motives ($F(1, 328) = .062$, $p = .803$, partial $\eta^2 < .001$). Therefore, H1b is supported while H2b is rejected.

H3 and H4 predicted that source-product incongruence and sponsorship disclosure respectively will result in greater consumer resistance. Significant main effects were observed for source-product congruence on consumer resistance ($F(1, 328) = 18.80$, $p < .001$, partial $\eta^2 = .054$). No main effect were found for sponsorship disclosure ($F(1, 328) = 1.64$, $p = .202$, partial $\eta^2 = .005$). The interaction effect between source-product congruence and sponsorship disclosure on consumer resistance was not present either ($F(1, 328) = .006$, $p = .936$, partial $\eta^2 < .001$). As such, results indicate that only source-product incongruence will result in significantly greater consumer resistance. H3 is supported and H4 is rejected.

H5 and H6 predicted that source product congruence and sponsorship non-disclosure will result in greater ad effectiveness. There is a significant main effect for source-product congruence ($F(1, 328) = 23.36$, $p < .001$, partial $\eta^2 = .066$). No main effect was found for sponsorship disclosure on ad effectiveness ($F(1, 328) = 1.03$, $p = .311$, partial $\eta^2 = .003$). There is no interaction effect between source-product congruence and sponsorship disclosure on ad effectiveness either ($F(1, 328) = .463$, $p = .497$, partial $\eta^2 = .001$). As such, the results indicate that source-product congruence will create more positive product and ad evaluations than source-product incongruence while sponsorship disclosure has no significant effect on ad effectiveness. H5 is supported while H6 is rejected. Table 1 summarizes all our hypotheses testing results.

[Table 1 about here.]

**Hierarchical Regressions on Consumer Resistance and Ad Effectiveness**

After finding out the positive main effects of source-product congruence on intrinsic motives, consumer resistance and ad effectiveness, further analysis was conducted to examine possible causal steps. In the online experiment, participants were asked to first indicate their perception of source motives before indicating their resistance toward the advertisement stimuli, and lastly to rate the effectiveness of the advertisement. This questionnaire sequence is in accordance with the stepwise process used in the hierarchical regression. A four stage hierarchical regression was conducted with consumer resistance and ad effectiveness as the dependent variables respectively. Variables such as consumer involvement, need for cognition, advertising skepticism and perceived sponsorship appropriateness were entered at stage 1 to control for individual differences. Source-product congruence was entered at the second stage and perception of source's intrinsic motives were entered at stage 3. For the regression on ad effectiveness, consumer resistance was entered in an additional stage 4. Regression statistics are reported in Table 2.

[Table 2 about here.]

The results revealed that in stage 1, consumer involvement ($\beta = 0.50$, $p < .001$) and advertising skepticism ($\beta = -0.12$, $p < .001$) contributed significantly to the regression model, accounting for 53.1% of the variance in ad effectiveness. Source-product congruence ($\beta = 0.21$, $p < .001$), introduced at stage 2, was also a significant predictor and introduced an additional 3.1% of variance. At stage 3, the source motives variables explained an additional 1.7% of variation, and both extrinsic ($\beta = -0.08$, $p < .05$) and intrinsic motives ($\beta = 0.13$, $p < .001$) were significant predictors. As the effect of source-product congruence became slightly weaker after including the perception of source intrinsic and extrinsic motives, we conclude that both motives partially mediated the relationship between source-product congruence and ad effectiveness. Lastly, consumer resistance ($\beta = -0.22$, $p < .001$), introduced at stage 4, was also a significant predictor accounting for an unique 5.7% of variance. As the significant relationships between perceptions of extrinsic ($\beta = -0.03$, $p > .05$) or intrinsic ($\beta = 0.06$, $p > .05$) motives and ad effectiveness in the previous step vanished, we conclude that consumer resistance fully mediated the relationship between source motives and ad effectiveness. Moreover, the effect of source-product congruence further decreased from step 3, we conclude that consumer resistance partially mediated the relationship between source-product congruence and ad effectiveness.

**Discussion and Conclusion**

Our results suggest that source-product congruence has a significant main effect on intrinsic motives. Compared to incongruent source-product match-up, source-product congruence comes across as more altruistic in persuasive intent, presenting a highly plausible case that the recommendation is motivated by an authentic sharing of product attributes. Meanwhile, we found that sponsorship disclosure has a significant main effect on extrinsic motive. Compared to not disclosing the sponsorship, sponsorship disclosure makes people feel stronger that the recommendation is motived by commercial and image-building intention. Both findings confirm that recognition of a persuasion item, through either source-product incongruence or sponsorship disclosure, activates people's persuasion knowledge. One piece of such knowledge is the attribution, which leads to their different perceptions of the SMIs' motives.

We also found that the discounting principle under the attribution theory and the change of meaning principle under the persuasion knowledge model (Yoon et al., 2006; Sørum, Grape, & Silvera, 2003) are at least partially validated. In particular, perceiving extrinsic motives increases while perceiving intrinsic motives deceases consumer resistance. When people detect that the SMIs are motived by money or image-building, they find the meaning of the persuasion message changes from authentic sharing to commercial promotion, and tend to discount the persuasion content they are exposed to by contesting or refuting the information via summoning their existing knowledge about persuasion.

In addition, we found that there is a multi-stage process (see Figure 1) where consumer's perception of source's motives mediates the impact of source-product congruence on consumer resistance and consumer resistance consequently mediates the relationship between congruence/motives and ad effectiveness. Our regression findings suggest that the mediation of motives and consumer resistance on the impact of congruence is partial, while the mediation of consumer resistance on the impact of motives is full. In other words, the extent to which people think the ad is effective is directly triggered by how much they challenge cognitively and emotionally the information they see in the persuasion content. Motives influence ad evaluations fully through activating/deactivating consumer resistance.

[Figure 1 about here.]

Also, our results indicate that perceived fit between the product and source of the sponsored content has a higher impact on consumer resistance and ad evaluations as compared to the presence of sponsorship disclosure. Contrary to previous studies which found a stepwise process where sponsorship disclosures result in recognition of the advertising that consequently generates distrust and lower intention to engage in eWOM (Boerman, Willemsen & Van Der Aa, 2017), our findings indicated that the disclosure merely activated consumer's conceptual persuasion and a perception of extrinsic motives, but did not trigger their subsequent attitudinal persuasion knowledge. This finding suggests that while current disclosure guidelines as stated by the FTC and ASAS successfully prohibit unfair persuasion with influencer marketing and allows for consumers to discern between posts with commercial intent, the commercial nature of Instagram posts is unlikely to affect advertising outcomes other than higher perceived extrinsic motives. This finding provides more understanding of consumer's behaviour on the Instagram platform that may be useful for both policy makers and marketers.

The nature of Instagram advertisements and the exposure duration towards the persuasive content may also have affected consumer's persuasion knowledge. According to the limited capacity theory (Lang, 2000), individuals require cognitive effort to process each message and are therefore unable to process all information simultaneously with limited cognitive resources (Buijzen, Reijmersdal & Owen, 2010; Lang, 2000). Boerman, van Reijmersdal and Neijens's (2012) study on the effect of duration of sponsorship disclosure on persuasion knowledge and brand responses found that although no difference was found in conceptual persuasion knowledge between participants who were exposed to the 3 and 6-second disclosure, a 6-second disclosure resulted in greater critical attitudes than a 3-second disclosure. Messaging processing theories also maintains that high levels of attention and processing are required for the activation of persuasion knowledge (Buijzen, Reijmersdal &

Owen, 2010; Petty, Ostrom, & Brock, 1981). Kahle & Homer (1985) suggests that readers of magazines advertisements, who have little involvement with the product class, completes their advertising evaluations relatively quickly. Similarly, the visual nature of Instagram posts facilitates quick scanning of communication content. In this case, it is possible that audiences have not processed the content elaborately enough. Further studies can examine the impact of exposure duration on the activation of both conceptual and attitudinal persuasion knowledge for source-product congruence and sponsorship disclosures on social media platforms such as Instagram.

   The present study consists of a few limitations, which presents opportunities for future research directions. Firstly, this study does not take into account possible differences that may arise from various degrees of incongruity with the adoption of a dichotomous operationalization of congruity (congruent vs. incongruent). Future research can consider looking into specific effects that different degrees of congruity would produce. By applying a non-dichotomous operationalisation, studies can further explore the non-monotonic relationship between congruence of source's perceived image and the endorsed product on consumer's resistance and evaluations. This experiment adopted two products of differing attributes – protein powder and a governmental finance website – to be paired with two respective expertise, nutrition and finance to examine the impact of source-product congruence. Although the two products showed relatively consistent findings, this study is limited to the two products adopted. For instance, exploration of how various product attributes such as associated financial or performance risks (Grewal, Gotlieb & Marmostein; 1994) may result in greater detail and specificity. Another limitation of this study pertains to the time horizon. Participant's reports of attitudes and behaviour were recorded immediately subsequent to the exposure to the advertisement. It is acknowledged that consumer responses to advertising may develop and even change over time after exposure to the advertisement (Obermiller, Spangenberg & MacLachlan, 2005). Over time, people may eventually start believing what they initially regarded as unbelievable (Gilbert, Krul & Malone, 1990).

Lastly, the participants recruited for this study were relatively young, with a mean age of 22.9 years old. While this age group makes up a large majority of Instagram users (61%) in Country A, younger online users who have been heavily exposed to social media throughout their lives may tend to be more trusting and tolerant of sponsored Instagram posts. Future studies using large samples of general Instagram users may help generalise the findings of this study to the general population.

In conclusion, this study adopts and extends the persuasion knowledge model and attribution theory in the examination of sponsorship disclosure and the match-up hypothesis in the context of influencer marketing in Country A's social media environment. An interesting discovery was that while disclosures successfully triggers consumer's persuasion knowledge, allowing for consumers to understand the commercial intent behind a sponsored post, it turns out that audiences are still able to negotiate the commercial intent and independently judge the authenticity of the message itself. As such, the fact that a post is known to be paid-for does not completely negate the expertise and perceived sincerity of the persuasive message. On the other hand, source product congruence has significant impacts on source motives, consumer resistance and perceived ad effectiveness. Careful consideration of source expertise and their applicability to the company's products must thus be taken to ensure that messages are effectively communicated. Results of the study lend support to the proposition that influencer endorsement is not unequivocally effective. Instead, the overall communicative effectiveness of influencer marketing is evaluated on a situational basis and one such situational factor is the alignment of source expertise to product attributes.

**Table 1.** Summary of Hypothesis Testing Results

| Hypothesis | IV | DV | Results |
|---|---|---|---|
| H1a | Source-Product Congruence | Extrinsic Motives | Rejected |
| H1b | Source-Product Congruence | Intrinsic Motives | **Supported** |
| H2a | Sponsorship Disclosure | Extrinsic Motives | **Supported** |
| H2b | Sponsorship Disclosure | Intrinsic Motives | Rejected |
| H3 | Source-Product Congruence | Consumer Resistance | **Supported** |
| H4 | Sponsorship Disclosure | Consumer Resistance | Rejected |
| H5 | Source-Product Congruence | Ad Effectiveness | **Supported** |
| H6 | Sponsorship Disclosure | Ad Effectiveness | Rejected |

**Table 2.** Hierarchical Regression Results Predicting Consumer Resistance and Ad Effectiveness.

| Variable | β | $T$ | sr² | $R^2$ | $\Delta R^2$ |
|---|---|---|---|---|---|
| Step 1: | | | | .53 | .53*** |
| Consumer Involvement | .50 | 16.60*** | .62 | | |
| Consumer NFC | -.01 | -.24 | -.01 | | |
| Advertising Skepticism | -.12 | -3.46*** | -.13 | | |
| Sponsorship Appropriateness | .06 | 1.79 | .07 | | |
| Step 2: | | | | .56 | .03*** |
| Consumer Involvement | .49 | 16.35*** | .60 | | |
| Consumer NFC | .01 | .21 | .01 | | |
| Advertising Skepticism | -.11 | 3.03** | -.11 | | |
| Sponsorship Appropriateness | .06 | 1.88 | .07 | | |
| Source-product Congruence | .21 | 4.84*** | .18 | | |
| Step 3: | | | | .58 | .02*** |
| Consumer Involvement | .45 | 14.04*** | .50 | | |
| Consumer NFC | .00 | .01 | .00 | | |
| Advertising Skepticism | -.09 | 2.72** | -.10 | | |
| Sponsorship Appropriateness | .05 | 1.58 | .06 | | |
| Source-product Congruence | .19 | 4.28*** | .15 | | |
| Extrinsic Motives | -.08 | -2.07* | -.07 | | |
| Intrinsic Motives | .13 | 3.39*** | .12 | | |
| Step 4: | | | | .64 | .06*** |
| Consumer Involvement | .38 | 12.09*** | .40 | | |
| Consumer NFC | .01 | 0.35 | .01 | | |
| Advertising Skepticism | -.12 | 3.81*** | -.13 | | |
| Sponsorship Appropriateness | .02 | 0.55 | .02 | | |
| Source-product Congruence | .13 | 3.09** | .10 | | |
| Extrinsic Motives | -.03 | -.74 | -.02 | | |
| Intrinsic Motives | .06 | 1.64 | .05 | | |
| Consumer Resistance | -.22 | -7.16*** | -.24 | | |

*Note.* $N = 336$; *$p < .05$, **$p < .01$, ***$p < .001$

**Figure 1**: Combined Results from ANCOVA and Hierarchical Regression

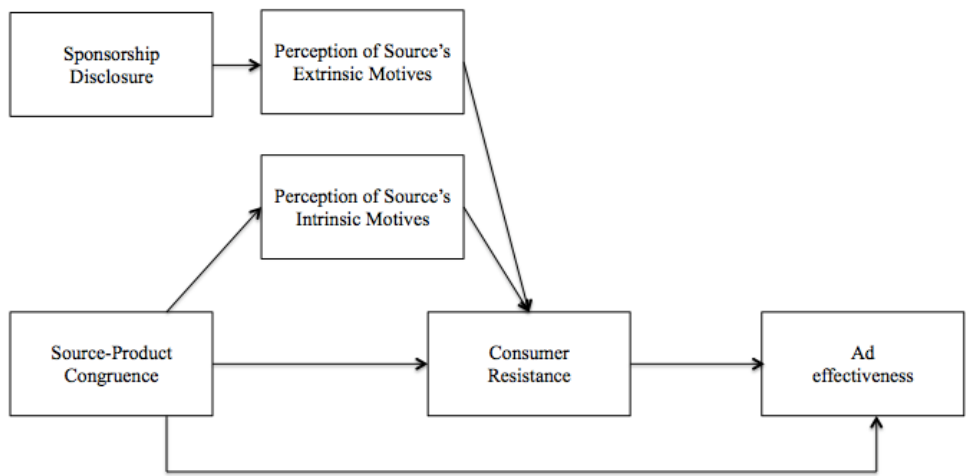